\documentclass[12pt]{article}
\usepackage{amsmath}
\usepackage[square,numbers,sort&compress]{natbib}
\usepackage{hyperref}
\usepackage{amssymb}
\usepackage{graphicx}
\usepackage{subfigure}
\usepackage{epstopdf}
\usepackage{multirow}
\usepackage{setspace}

\doublespace
\begin{document}
\begin{center}
{\Large\bf Forced-rupture of Cell-Adhesion Complexes Reveals abrupt switch between two Brittle States}\\
\ \\
{\large Ngo Minh Toan$^1$ and D. Thirumalai$^{2}$}\\
\ \\
$^1$  Biophysics Program, Institute for Physical Science and Technology, University of Maryland, College Park, MD 20742\\
$^{2}$  Department of Chemistry, The University of Texas at Austin, TX 78712
\end{center}
\newpage
\begin{abstract}
Cell adhesion complexes (CACs), which are activated by ligand binding, play key roles in many cellular functions ranging from cell cycle regulation to mediation of cell extracellular matrix adhesion.  Inspired by single molecule pulling experiments using atomic force spectroscopy on leukocyte function-associated antigen-1 (LFA-1), expressed in T-cells, bound to intercellular adhesion molecules (ICAM), we performed constant loading rate ($r_f$) and constant force ($F$) simulations using the Self-Organized Polymer (SOP) model to describe the mechanism of ligand rupture from CACs. The simulations  reproduce the major experimental finding on the kinetics of the rupture process, namely, the dependence of the most probable rupture forces ($f^*$s) on $\ln r_f$ ($r_f$ is the loading rate) exhibits two distinct linear regimes. The first, at low $r_f$, has a shallow slope whereas the the slope at high $r_f$ is much larger, especially for LFA-1/ICAM-1 complex with the transition between the two occurring over a narrow $r_f$ range.  Locations of the two transition states (TSs), extracted from the simulations show an abrupt change from a high value at low $r_f$ or $F$ to a low value at high $r_f$ or $F$. This unusual behavior in which the CACs switch from one brittle (TS position is a constant over a range of forces)  state to another brittle state is  not found in forced-rupture in other protein complexes.  We explain this novel behavior by constructing the free energy profiles, $F(\Lambda)$s, as a function of a collective reaction coordinate ($\Lambda$), involving many key charged residues and a critical metal ion ($Mg^{2+}$). The TS positions in F($\Lambda$), which quantitatively agree with the parameters extracted using the Bell-Evans model, change abruptly at a critical force, demonstrating that it, rather than the molecular extension is a good reaction coordinate.  Our combined analyses  using simulations performed in both the  pulling modes (constant $r_f$ and force) reveal a new  mechanism for the two loading regimes observed in the rupture kinetics in CACs.
\end{abstract}
\date{today}

\section*{Introduction}
Ligand-receptor interactions govern a number of cellular functions.   Amongst these are cell adhesion complexes (for example ligands bound to integrins{~\cite{Hynes02Cell}, selectins~\cite{Somers00Cell}, and cadherins~\cite{Weber11JCellScience}),  which perform multiple functions, such as anchoring, migration, signaling and division. Integrins belong to a family of proteins that play crucial roles both in cell mechanical support (adhesion) as well as in signal transduction, especially in immune system's activities, such as leukocyte trafficking~\cite{SpringerReview2007}. Single molecule pulling experiments of Leukocyte function-associated antigen-1 (LFA-1) with intercelluar adhesion molecules (ICAM-$x$ with $x$ = 1,2, or 3) have provided glimpses of interactions stabilizing these complexes. The forces at which the rupture of the ligand (LFA-1) occurs, when the complex is stretched, is an indication of the stability of the cell adhesion complexes (CAC). Based on single molecule force spectroscopy, using atomic force microscopy (AFM), LFA-1/ICAM-$x$ ($x$ = 1 and 2), with pulling speeds spanning three orders of magnitude, Moy and coworkers~\cite{Experiment2}  suggested that LFA-1 ruptures fully first by crossing an outer and then an inner barrier. Such an interpretation is reminiscent of a similar picture proposed by Merkel wt. al.,~\cite{Merkel99Nature} in a pioneering study involving biotin or avidin in complex with streptavidin. This interpretation was proposed because the [$f^*, \ln r_f$] plot, a graph of the most probable rupture force, $f^*$,  as a function of $\ln r_f$ with $r_f$ being the loading rate, exhibits two distinct linear regimes.

%Because single molecule experiments have access to only one reaction coordinate, the molecular extension of the complex that is conjugate to the applied force, the details of the rupture process cannot be easily determined by experiments alone. Several previous studies in a variety of systems (proteins, RNA, viruses, and complexes) have established that simulations based on the coarse-grained Self-Organized Polymer (SOP) model
%Thankfully,  there are many cases in which the marriage between experiments and simulations, especially derived from the coarse-grained Self-Organized Polymer (SOP) models, 
%have provided a molecular bases of their functions. interpret~\cite{RiefDTGFP,Kononova2014JACS,zhmurov2011Structure,Dima10BJ,Bauer15PNAS,Zhuravlev14JMB,Chen09PNAS,Hori17JMB}. The availability of experimental data on mechanical unbinding of the LFA-1/ICAM-$x$ complexes\cite{Experiment2}, has prompted us to investigate the effect of force on CACs using SOP simulations in order to elucidate in detail the rupture mechanism. % In particular,  we hope to extensively explore the systems in this aspects through simulations.

In order to provide quantitative insights into the adhesion of the leukocyte cells to cell adhesion matrix {\em en route} to inflammation sites, which is facilitated through the integrin-ICAM complexes, Moy et al. performed a set of single molecule pulling experiments~\cite{Experiment2} using atomic force microscopy (AFM). They measured the dynamic force spectra (DFS) to determine the interaction between leukocyte function-associated antigen-1 (LFA-1, or the integrin $\alpha$L$\beta$2) and intercellular adhesion molecule-1 and -2 (ICAM-1 and ICAM-2) at the single-molecule level. The experiments were performed by attaching a Jurkat cell to the tip of an AFM cantilever (see Fig. 1 in~\cite{Experiment2}) and then allowed to contact the LFA-1 molecules on the cell surface with ICAM-1 or ICAM-2 molecules on a substrate. This procedure resulted mostly in single LFA-1/ICAM-1(-2) bonds. The mechanical rupture events were recorded by pulling the cantilever at a constant speed spanning three orders of magnitude. At this level, the experiments involved the whole system of cell-integrin-ICAM-1(-2), while the binding of the integrin to the ligands only occurred at the interface between the I domain, which is a single domain in more than fifteen domains of the integrin, and the D1 domain of the ICAM-1(-2)~\cite{EvansRev2001,Franz2007,Experiment2,Taubenberger07MolBiolCell} (see Fig.~\ref{fig:ID1}). The analyses of the [$f^*, \ln r_f$] plot using the Bell-Evans model~\cite{Bell78Science,Evans1997} showed two loading regimes.   Two-step rupture  has been explained using a one-dimensional free energy landscape with two barriers~\cite{Merkel99Nature,Experiment2}. The theoretical justification for postulating the two-barrier picture was given much later~\cite{Hyeon12JCP}. However, the mechanism underlying the abrupt change in the [$f^*, \ln r_f$] plot in the CACs, as opposed to the more gradual change observed previously~\cite{Merkel99Nature} requires investigation. 
%further slope of unfolding rate or the most probable force as a function of the loading rate is not understood.  The free energy profiles with two barriers  of the complexes needed to explain the data could, therefore, be viewed as  a speculation.

Because single molecule experiments have access to only one reaction coordinate, the molecular extension of the complex that is conjugate to the applied force, the details of the rupture process cannot be easily determined by experiments alone. Here, we used the SOP model~\cite{Hyeon06Structure} with modifications for CACs and Brownian Dynamics to investigate the rupture dynamics in the mechanical unbinding of LFA-1 from ICAM-1 and ICAM-3. Several previous studies~\cite{RiefDTGFP,Kononova2014JACS,zhmurov2011Structure,Dima10BJ,Bauer15PNAS,Zhuravlev14JMB,Chen10PNAS,Hori17JMB}. in a variety of systems (proteins, RNA, viruses, and complexes) have established that simulations based on the coarse-grained Self-Organized Polymer (SOP) model
%Thankfully,  there are many cases in which the marriage between experiments and simulations, especially derived from the coarse-grained Self-Organized Polymer (SOP) models, 
have provided a molecular bases for interpreting and predicting the outcomes of experiments.   The availability of experimental data on mechanical unbinding of the LFA-1/ICAM-$x$ complexes\cite{Experiment2}, has prompted us to investigate the effect of force on CACs using SOP simulations in order to elucidate in detail the rupture mechanism.  In order to obtain reliable statistics in the simulations, we retained only the essential domains of the complex, namely the $\alpha$L I domain and domain D1 of ICAM-1/3 shown in Fig.~\ref{fig:ID1}. Due to the unavailability of the crystal structures of the LFA-1/ICAM-2 complex in the Protein Data Bank (PDB), we did not perform a similar study on this complex.

To directly compare our data to the corresponding experimental studies, we first employ constant loading rate or force ramp mode for the mechanical unbinding. The simplicity of the SOP model allowed us to perform simulations over a wide range of loading rates spanning five decades. The smallest loading rates in the simulations  overlap with those used in the experiments. There are several key results in this paper. (i) To our satisfaction, the simulated [$f^*, \ln r_f$] plot displayed a fast loading regime and a slow loading regime as reported in the AFM pulling experiments~\cite{Experiment2}. More importantly, the extracted kinetic parameters from the simulations  agree  well with those obtained based on the experimental data. (ii) We show that the transition state (TS) location jumps abruptly between two values as $r_f$ is increased, in accord with experiment (see Fig. 6 in~\cite{Experiment2}).  Although the two-barrier picture was inferred using the AFM data, our  work is the first to demonstrate it using simulations.  Using a structural interpretation, we propose that CACs switch from one brittle (TS location is insensitive to force~\cite{HyeonJPCM2007}) state to another over a narrow range of $r_f$ or constant force ($F$).  It is worth emphasizing that no parameter in the SOP energy function was adjusted to obtain agreement with experiments, which not only attests to the success of the SOP model but also shows the response of biological molecules to force is encoded in the topology~\cite{Klimov00PNAS}.  (iii) Because a molecular extension with a single or two barriers is not an appropriate reaction coordinate, we  constructed collective coordinate, $\Lambda$ based on the structure of the complex,  in terms of which the $F$-dependent free energy profiles ($F(\Lambda)$s) have a single barrier. Surprisingly, the TS location in $F(\Lambda)$ jumps abruptly at a critical force, further showing that CAC states are brittle.  
%Although this is only of theoretical interest such profiles might provide insights into the ductility of the CACs.

\section*{Results}
\subsection*{Methods}
{\it Model:} In the experiments~\cite{Experiment2}, force spectroscopy was used to study the unbinding of LFA-1 ($\alpha$L$\beta$2) from ICAM-1 and ICAM-2. Here, using the SOP model with modifications (described in the Appendix) and the structures available in the Protein Data Bank (PDB), we study the mechanical unbinding of the complexes between LFA-1 and ICAM-1 and ICAM-3. To perform simulations with good statistics, we use the SOP model with only the $\alpha$L I domain and the binding domain in the ligand molecules, namely the D1 domain of ICAM-3. In the simulations, the PDB entries 1MQ8 and 1T0P~\cite{GangSong2005} serve as the starting structures. Several residues of the D1 domain are tethered with springs to an immobilized substrate mimicking the procedure used in the AFM experiments~\cite{Experiment2}. On the opposite side of the complex, several residues of the I domain are tethered to an imaginary plane, which in turn is pulled with a spring, either at a constant speed $v$ or constant force ($F$), to mimic the action of the AFM cantilever. The extension is defined as the distance from the average position of the tethered I domain residues to that of the tethered D1 domain residues (seeAppendix for additional details).

{\it Theory:} We analyze the data using the Bell model~\cite{Bell78Science} and the Evans-Ritchie theory~\cite{Evans1997}, as was done in the experiments~\cite{Experiment2}. The off rate of a ligand-receptor complex subject to force $f$ is  related to that in the absence of force $k^o$ as, $k(f) =  k^o \exp\left({f\gamma\over k_BT}\right)$ with $\gamma$ being the position of the the transition state, assuming that the pulling coordinate is a reasonable choice of the reaction coordinate. At a constant loading rate $r_f = k_{AFM}v$ ($k_{AFM}$ is the stiffness associated with the cantilever), the probability distribution for the unbinding of the complex is~\cite{Evans1997,Experiment2},
\begin{eqnarray}\label{eqn:p(f)}
P(f) &=& k^o \exp\left({f\gamma\over k_BT}\right)\exp\left({k^ok_BT\over \gamma r_f}\left[1 -\exp\left({f\gamma\over k_BT}\right)\right]\right)
\end{eqnarray}
which gives the most probable unbinding force $f^*$ as
\begin{eqnarray}\label{eqn:f*}
f^* & = & {k_BT\over \gamma}\ln\left({\gamma\over k^ok_BT}\right) + {k_BT\over \gamma}\ln(r_f). 
\end{eqnarray}
Note that $f^*$ varies linearly with $\ln r_f$. In this standard formulation there is only one transition state, whose position does not change with $r_f$. 

In the distribution in Eq.~\eqref{eqn:p(f)}, subject to  the condition that it is to be justified {\it a posteriori} $r_f\gg k^o{k_BT\over \gamma}$, the mean force $f_m$ is  related to the most probable force $f^*$ as $f_m  =  f^* - {k_BT\over \gamma}\gamma_E$, where $\gamma_E \approx 0.577$ is the Euler gamma constant. Thus, $f_m$ is also linear in $\ln r_f$ at large $r_f$. If the number of sampled forces is  small, it is more accurate to determine the mean than the most probable value, this equation means that we can obtain $f^*$ from $f_m$ with better accuracy.

\subsection*{Two-loading regimes in the DFS of $\alpha$L I domain/D1}
Fig.~\ref{fig:fpiezo} is a typical simulated unbinding trajectory of the  LFA-1/ICAM-3 complex when stretched at a constant $r_f$. In the graph, the force exerted on the imaginary plane (see Methods in the previous section) is plotted against the spring displacement - the analogue of the piezo displacement in experiments~\cite{Experiment2}. First, the force increases linearly with the displacement up to a point when there is a sudden drop to zero, which signals the unbinding of the complex. The drop is simultaneous with the sudden increase of the extension of the complex. Such a trace is similar to what is observed in unbinding experiments~\cite{Franz2007,Experiment2}. We fit the linear portion of the curve with a straight line to determine the effective spring constant $k_\text{trans}^\text{eff}$ and the unbinding force. We find that $k_\text{trans}^\text{eff} \approx 7.4$pN/nm, which is smaller than the nominal value $k_\text{trans} = 10$ pN/nm because of rotation effects. The behavior in Fig.~\ref{fig:fpiezo} is found in  all the unbinding trajectories at different pulling velocities  spanning five orders of magnitude. The effective spring constant together with the pulling velocity $v$ determines the loading rate $r_f = k_\text{AFM}^\text{eff}v$. We analyze the relation between the rupture force, $f$, and the loading rate to determine the distribution of the unbinding forces.

In Fig.~\ref{subfig:forcePDF} we plot the distribution of the rupture forces of the LFA-1/ICAM-1 complex at several representative loading rates calculated from  the simulations. The fit, using eq.~\eqref{eqn:p(f)} to the distributions, gives  the most probable unbinding forces, $f^*$s. Fig.~\ref{subfig:forcerf} show  the [$f^*, \ln r_f$] plots extracted from simulations for the simplified LFA-1/ICAM-1 or -3 complex.  In the same graph, we also show the [$f^*, \ln r_f$] plots  found in experiments for the full LFA-1/ICAM-1 and -2 complexes~\cite{Experiment2}. We immediately notice that although the simulated curves lie below the experimental curves, the rupture forces are well within the experimentally measured range.  In addition, the low loading rate portions of the simulated curves appear to converge to the experimental values.  The most striking feature in Fig.~\ref{subfig:forcerf} is that the simulated curves also exhibit two distinct linear regimes in the [$f^*, \ln r_f$] plots.    We conclude that the simulation results are consistent with experiments. The two linear regimes in the [$f^*,\ln r_fF$] plots could be interpreted to mean that that only by overcoming two different intermolecular potential barriers do the CACs rupture~\cite{Experiment2}. We discuss the meaning of the two loading regimes using free energy profiles as a function of a structure-based a one-dimensional scalar reaction coordinate in detail below.

Using equation~\eqref{eqn:f*}, we fit the  two linear portions of the simulated DFS curve. The fitted parameters values are given in Table 1. The indices denote different portions of the curve~\cite{Experiment2} with $1$ for the steeper slope, and $2$ for the shallow slope (see Fig.~\ref{subfig:forcerf}). Note that in the PDB structures, used as input in our simulations, the I domain is in an open high-affinity state. Therefore, it is appropriate to compare the extracted values with those for similar complexes in high-affinity state. In Table 1, we also cite the fit values for LFA-1/ICAM-1 and LFA-1/ICAM-2 complexes treated with Mg$^{2+}$/EGTA as in ref.~\cite{Experiment2} for  comparison. Most of the values obtained from SOP simulations are within a factor of two-three compared to values inferred from experiment. Moreover, it has been reported that $k_{off}$ of LFA-1/ICAM-3 complex is about three times larger than the value for LFA-1/ICAM-1 and ICAM-2 complexes~\cite{GangSong2005}. We expect that the discrepancies in $k^o$ amongst the three systems would be similar to those in $k_{off}$. Our values for $k^o$ of the ICAM-3 complex are, therefore, in good agreement with the reported values for LFA-1/ICAM-1 and ICAM-2 complexes. Given the coarse-grained nature of the SOP model and the fact that our systems are simplified compared to the the two systems studied in the experiments, the agreement should be considered to be very good. Clearly, we could improve the agreement by adjusting the SOP force-field parameters, which would negate the requirement of transferability of the coarse-grained model. We note, {\it en passant} that it is current impossible to obtain similar accuracy using atomic detailed simulations of CACs subject to force.

Next, the fits of Eq.~\eqref{eqn:p(f)} to the simulated distribution of unbinding forces of the LFA-1/ICAM-1 and -3 complexes, were used  to determine $\gamma$,   the locations of the transition states. The values of the TS positions as a function of the loading rate $r_f$ are plotted in Fig.~\ref{fig:gammarf}. It is transparent that there are two plateaus  corresponding to the two  values of $\gamma$, which do not change with $r_f$ in the low and high $r_f$ range. One of them is  in the range of  $r_f$ from $10^3$, and other covers the range $2\times10^5$ pN/s and $10^6$ to $3\times 10^7$ pN/s. The stable values are excellent agreement with fits to the [$f^*, \ln r_f$] plots in Fig.~\ref{subfig:forcerf} and listed in Table 1. The coincidance of the  ranges of loading rate over which the there is a shift in $\gamma$   with the ranges of the two loading regimes in Fig.~\ref{subfig:forcerf} further confirms the consistency of the data analyzed with two approaches to the same theory. In other words, no additional information about $\gamma$ is discernible by analyzing the  data in Fig.~\ref{subfig:forcePDF}.

It is surprising and indeed intriguing that there is an abrupt change in $\gamma$ versus $\ln r_f$ plot (Fig.~\ref{subfig:forcerf}) with no intermediate values between the two plateaus in the $2\times 10^5\text{pN/s}\le r_f\le 10^6\text{pN/s}$ range. In order to appreciate the claim of intrigue, it should be noted that the [$f^*,\ln r_f$] plot for biotin/avidin-streptavidin comples exhibit a more gradual change than what simulations and more importantly experiments show (especially for LFA-1/ICAM-1) for CACs (see Fig.~\ref{subfig:forcerf}). Could the abrupt change in $\gamma$ imply that there is only a single transition state, whose position changes from one value to another as the loading rate increases? Indeed, this is the case as we demonstrate below. 
%Such an interpretation result in a free energy profile drawn schematically in Figure 6 of~\cite{Experiment2}. 

\subsection*{Load-dependent switch between two brittle CAC states explained}
In order to better understand the underlying mechanism for the two-regime behavior of the DFS, we performed additional simulations at {\em constant} pulling force mode. The purpose is  to compute the free energy profile (FEP) of the intermolecular potential of the complex so that quantitative insights into the unexpected jump in the TS can be provided. The first task is to choose an appropriate reaction coordinate.  The molecular extension, $R$, of the complex, defined previously, could be a natural choice for the reaction coordinate. 

In order to explain the two loading regimes observed in experiments the free energy as a function of $R$ must have an outer and inner barrier. Such a profile does account for the large (small) value of $\gamma$ at small (large) force~\cite{Experiment2,Hyeon12JCP}. However, from such a profile we would predict that both the TS locations would change continuously by altering the applied force, which contradicts the findings in experiments and our simulations. Thus, an alternative explanation is required for the abrupt change in the value of $\gamma$ over a narrow range of $\ln r_f$. 

From the structural perspective of the CACs, we notice that the binding/unbinding transition is related to the pairs of extra-domain residues that directly interact with one another. Based on this observation, we constructed a collective  reaction coordinate as follows. A binding pair is defined as two residues, one in the I domain and the other in the D1 domain, that are within a cutoff radius (chosen as 8\AA\ in the SOP model~\cite{Hyeon06Structure}) in the ground state of the CAC. We also regard the coordinating residue in the D1 domain and $Mg^{2+}$ ion~Fig.\ref{ID1} as a binding pair because of the strong electrostatic attraction.  The attractive interactions between each binding pair  provide much of the stability between the two domains. When the force is high enough for all the binding pairs to reach beyond the ranges of attraction, the CAC is ruptured.  Therefore, we expect that a reaction coordinate, $\Lambda$, chosen as the simple arithmetic mean of the binding pair distances, would be suitable for our purpose.

To make sampling possible for accurate computations of the FEPs, we artificially make the system closed by introducing a potential that prevents the two domains from moving too far apart. The potential is so chosen that it does not alter the free energy profile in the relevant regions of interest. Furthermore, we choose the potential to ensure that the first derivative with respect to $\Lambda$ is continuous.   Such a  potential is constructed as follows. Whenever the reaction coordinate, $\Lambda$ (Eq. \ref{rxncor}) exceeds a critical value $\Lambda_c$, the potential is different from zero, and is added to the SOP potential $U_0$ of a binding pair, $i$and $j$. The form of the potential is,
\begin{eqnarray}
U(r_{ij}) & = & U_0(r_{ij}) + \left(e^{a(\Lambda-\Lambda_c)} - \left[1 + a(\Lambda-\Lambda_c)\right]\right)
\end{eqnarray}
where $a = 2$\AA$^{-1}$ is the stiffness of the exponential term, and
\begin{eqnarray}
\Lambda &= & {1\over N_{bd}}\sum_{\text{all binding pairs}} r_{i,j}
\label{rxncor}
\end{eqnarray}
with $N_{bd} = 29$ for the I domain/D1 complex. Because $U(r_{ij})$ is smooth at $\Lambda-\Lambda_c$ up to second order in $\Lambda-\Lambda_c$,  the derivative of $U(r_{ij})$ with respect to $\Lambda$ is continuous at $\Lambda-\Lambda_c$.

In Fig.~\ref{subfig:pdf} we plot the distribution, $P(\Lambda)$, of the reaction coordinate $\Lambda$ of the LFA-1/ICAM-3 complex for different {\em constant} pulling forces. We do not report results for the the LFA-1/ICAM-1 complex because the barrier in this case is considerably  higher, which prevents us from obtaining reliable statistics.   Interestingly, the distribution, $P(\Lambda)$, is bimodal with one peak at the ground state value of $\Lambda_{nat}$, and the other is at around $\Lambda_c$. This implies that when in terms of $\Lambda$ free energy landscape of the CAC is simple, implying that $\Lambda$ could be the ideal reaction coordinate for this complex.  Fig.~\ref{subfig:FEPs} shows plots of the free energy profiles, $F(\Lambda^{\prime}) = -k_BT\log P(\Lambda^{\prime})$ where $\Lambda^{\prime} = \Lambda - \Lambda_{nat}$, and $\Lambda_{nat}$ is the value of $\Lambda$ in the ground state of the CAC. 

All the curves in Fig.~\ref{subfig:FEPs} have been shifted horizontally so that the ground state (as found in the crystal structure) is at the origin, and vertically so that the free energy at the origin is zero. With this shifting, the intrinsic or unaltered part of the free energy profile spans the relevant region $\Lambda^{\prime} < \Lambda_c^{\prime} \equiv \Lambda_c - \Lambda_{nat}$. For different forces, we chose different values of $\Lambda_c$ so that all the curves could be compared on the same footing.  Note that due to the particular choice of reaction coordinate, any two curves are not simply related through a tilting term proportional to the difference in the force $-\Delta F \Lambda$. However, the larger the pulling force is, the more tilted to the right the curve becomes, which is qualitatively similar to profiles with the molecular extension as the reaction coordinate.

For each profile, the position of the transition state $\Lambda^{TS}$ is determined as the location of the  maximum in $F(\Lambda^{\prime})$ (Fig.~\ref{subfig:FEPs}), which satisfies  $0 < \Lambda^{\prime} \le \Lambda_c^{\prime}$. For forces $F = 60$ and $50$ pN, $\Lambda^{TS}_1$ is more or less localized  around $\Lambda^{TS}_2 = 7$\AA. On the other hand, at larger forces, $F = 70$ and $80$ pN, the transition state appears to stay around $\Lambda^{TS}_1 = 4$\AA. Interestingly, at an intermediate force, $F = 67$ pN, there is a plateau in the energy profile extending almost from $\Lambda^{TS}_1$ to $\Lambda^{TS}_2$. This suggests that $F = 67$ pN is the critical force at which the transition state position abruptly changes from $\Lambda^{TS}_2$ to $\Lambda^{TS}_1$. Recalling that the Bell model parameters obtained from fitting the DFS curves (Table 1), $\gamma_1 = 3.8 \pm 0.5$\AA\ and $\gamma_2 = 7.8 \pm 0.7$\AA, we see that $\Lambda^{TS}_1$ is almost identical to $\gamma_1$ and  $\Lambda^{TS}_2  \sim \gamma_2$. Moreover, the critical force separating the slow and fast loading regimes for LFA-1/ICAM-3 (solod blue square) in Fig.~\ref{fig:forcerf} for the DFS is around 67 pN, value that is the same in the equilibrium $F(\Lambda)$ (Fig.~\ref{subfig:FEPs}) and the crossover force in the [$f^*,\ln r_f$] plot (Fig~\ref{fig:forcerf}b). Thus, the  predicted jump in $\gamma$ at constant force equilibrium free energy profiles  agrees  with the analysis based on the [$f^*, \ln r_f$] plot using the Bell-Evans theory. %They mutually confirm each other, and confirms that that the collective variable $\Lambda$ is a good reaction coordinate in contrast to the molecular extension.

A consistent picture has emerged from analyzing both $F(\Lambda)$ and [$f^*, \ln r_f$] plots.  What is interesting in this key finding is that we have uncovered  a new mechanism for obtaining two loading regimes (discussed  below). It is natural to think of an energy profile which has two local maxima and two local minima, one corresponding to the ground state and the other is at some position in between the two maxima to explain the observed two-regime behavior of the DFS~\cite{Experiment2}. Indeed, the data in~\cite{Experiment2} could be explained using  a one dimensional free energy profile with an inner and an outer barrier, as first shown in~\cite{Merkel99Nature}, and justified subsequently more recently~\cite{Hyeon12JCP}. In such a scenario, the TS location would not change abruptly as our simulations demonstrate, which also supports the analysis based on the experimental [$f^*, \ln r_f$] plots.   However, using the reaction coordinate $\Lambda$, we have shown that switches in the TS position with increasing loading rate or increasing constant force occur (Fig.~\ref{fig:pdf} (b)) abruptly over a narrow range of forces. Our interpretation of the abrupt TS switch is  that the CACs transition between two brittle states, whereas the two-barrier picture would suggest the CACs are plastic (TSs move with force) in their response to mechanical forces. 
% without the second local minimum in between the two local maxima, hence without the first local maximum, an energy profile having the properties as seen in Fig.~\ref{fig:pdf} (b) in our model can also lead to the same intriguing behavior.

\subsection*{Structural Insights}
To better understand the proposed physical picture from a structural perspective, we take a closer look at the potentials between the binding pairs used in defining $\Lambda$ in Eq. \ref{rxncor}. In Fig.~\ref{fig:fep} (a), (b) and (c) we plot the FEPs for the individual binding pairs at 50, 67 and 80pN, respectively. To improve visualization, the individual profiles have been shifted horizontally by the distances in the ground state of the CAC. For comparison  the  $F(\Lambda)$s at the three forces are also plotted on the same graphs. The curves can be roughly be divided into three groups, each with roughly the same number of profiles: strongest binding, intermediate binding and weakest binding ones. At forces lower than 67pN, the weakest group profiles have not been tilted by the force completely, and still exhibit a local minima at the distances in the ground state. At 67pN, the profiles are almost flat. Above this force, they are completely tilted and the intermediate group profiles become flat.  

Fig.~\ref{fig:fep} (d) shows the residue numbers of the binding pairs in corresponding colors. It can be seen that group 1 mostly includes residue E37 of the D1 domain {\em and} one of the residues from D140 to S142 of the I domain. Group 2 includes residues that are either the nearest neighbor or next nearest neighbor of E37 {\em and} residues from D140 to S142 on the I domain. In group 3, the residues are far away from these residues. E37, D140 and S142 are 3 of the 6 residues that are coordinated to the $Mg^{2+}$ ion, and are dominated by electrostatic interactions.  Consequently,  the interactions involving  these amino acid residues  are stronger than those involving other pairs. Taken together these results suggest that at forces lower than 67pN,  the binding pairs in groups 2 and 3  rupture simultaneously, while at higher forces, the strongest group is the last to be ruptured. These two events explain in structural terms the two brittle states, which manifest themselves as a sudden change in the transition states. 

\section*{Discussion and Conclusions}
Here, we have reported the results of our study, based on  simulations probing the forced-rupture of ligands from cell adhesion molecules. The good agreement between simulations and experiments  once again confirms the reliability of the SOP model in predicting and interpreting the unfolding pathways of proteins and their complexes.  The simulations, performed under conditions that are close to the experimental loading rates, provide the much-needed molecular interpretation of the mechanism of forced-rupture of these important class of protein complexes. Only by combining experiments and thoughtful computations can the power of single molecule experiments be greatly enhanced, as we recently showed in our theoretical study of parallel unfolding pathways in small single domain proteins~\cite{Zhuravlev16PNAS}.

We conclude with the following additional remarks. 

\begin{itemize}
\item

The response of cell adhesion complexes to mechanical forces is diverse, and different scenarios are required to infer the rupture mechanisms using experimental data.  (i) In some instances, for example antibodies, G1 and DREG56, bound to P-selectin and L-selectin~\cite{Marshall03Nature,Sarangapani04JBC}, respectively, the logarithm of the lifetimes of the CACs decrease linearly with force~\cite{Barsegov06JPCB}, as predicted by the classic Bell model~\cite{Bell78Science}. This situation corresponds to the so-called slip bonds.  (ii) In an  important study on the rupture of ligands from the CAC (P-selectin)  Zhu and coworkers~\cite{Marshall03Nature} demonstrated the counter intuitive force response in which the lifetimes first increase (catch bonds).   Only at higher forces the life times decrease, as expected for slip bonds. Interestingly, catch bonds have been recently found in a number of protein-ligand complexes that are unrelated to CACs~\cite{Huang17Science,Buckley14Science,Luca17Science}. Phenomenological theories, based on two state models~\cite{Barsegov05PNAS,Thomas2008ARB}, and microscopic theory~\cite{Chakrabarti17JSB,Chakrabarti14PNAS} have been proposed to explain the catch bond behavior at low forces and subsequent transition to slip bonds at high forces.    In cases when the Bell model or a variant is applicable~\cite{Dudko06PRL} the reaction coordinate of choice is  the experimentally accessible extension ($R$), conjugate to the applied force. (iii) In the example of biotin or avidin bound to streptavidin [$f^*,\ln r_f$] plot shows curvature, which is tidily explained using a one dimensional  free energy profile containing  two barriers with $R$ being the reaction coordinate.   The movement of the location of the transition state, in such a one dimensional free energy profile, is an indication of the ductility of the complex. Brittle behavior implies that the TS does not change as force changes and plasticity means that the TS moves as the value of force is altered. We should note that it is difficult to construct physically reasonable $f(R)$ with a single barrier~\cite{Hyeon12JCP} to describe ligand unbinding in biotin-streptavidin complex although in forced-unfolding of monomeric proteins a range of mechanical behavior may be described using such a picture~\cite{Cossio16BJ}.

In contrast to the scenarios described above, CACs complex exhibit two distinct slopes in the loading-rate dependent unbinding force. The change in the slope is very pronounced in CACs, especially LFA-1in complex with ICAM-1. The logical interpretation is that the underlying free energy profile must be described by two activation barriers with the outer (inner) barrier being important at low (high) forces~\cite{Merkel99Nature,Experiment2,Hyeon12JCP}. This would be not be inconsistent with our simulations as well. However, the switch from a large value corresponding to  the outer TS location to a small value corresponding to the inner barrier is abrupt, which is not expected if the TS moves continuously, as seems to be the case in the forced-unbinding of biotin from streptavidin. Our findings here imply that there are two brittle transition states in the CACs, one at high forces and the other at low forces. It is unclear if the novel mechanism discovered here  is a consequence of using high forces both in AFM experiments as well as simulations. It would be most instructive to perform similar experiments using optical tweezer experiments. It would be interesting to assess if this new mechanism of ligand rupture,  predicting  sharp changes in the TS location, is applicable to other protein complexes as well. 

\item
Only when the logarithm of the unbinding rate ($\ln k_u$) is linear in the applied force or the  [$f^*, \ln r_f$] plot  is linear, can one surmise that molecular extension is a reasonably good reaction coordinate. This is certainly not the case for the rupture of CACs by force. The abrupt change in the TS location cannot be explained by TS movement in a one dimensional free energy profile with extension as the reaction coordinate or by parallel rupture pathways, which manifests itself as an upward curvature in the force-dependence of the $\ln k_u$~\cite{Zhuravlev16PNAS}. Surprisingly, in CACs we uncovered a complex collective reaction coordinate (Eq. \ref{rxncor}) in which the free energy profiles exhibit a simple behavior. The intermolecular free energy profile as a function of $\Lambda$ in Eq. \ref{rxncor} shows an abrupt jump in the location of the TS.  Using $\Lambda$ as the reaction coordinate the response of CACs to force can be described using a two-state model. Whether uncovering such coordinates is possible for arbitrary complexes is unclear. Nevertheless, we can conclude that responses of protein complexes are likely to be varied, and could well depend on the context in which they function.    

\end{itemize}

\section*{Appendix}
In the SOP model~\cite{Hyeon06Structure} each amino acid in the CAC complex is represented by a singe interaction center located at the $C_{\alpha}$ atom.   The values of the few parameters characterizing the energy function in the SOP model are given elsewhere \cite{Hyeon06Structure} %kept as in the previous one (cite Idomain) 
The cantilever spring constant was chosen to be $k_\text{trans} = 10.0$ pN/nm similar to the value of the AFM cantilever in the unbinding experiments~\cite{Experiment2}. The major advantage of the SOP model is that forced unfolding simulations can be performed closely mimicking the conditions used in experiments, which is not possible in atomic detailed simulations.  To investigate the rupture of the CACs triggered by mechanical forces, we introduce modifications to the SOP model, the details of which are given below. 

The I domain/D1 complex is treated as a single protein in the SOP model except that there is no backbone connectivity between the I domain and the D1 domain. The interactions of the $Mg^{2+}$ ion with the five residues in the I domain and one residue in the D1 domain are modeled as follows. Each interaction energy consists of two terms, one is the excluded volume interaction. The  functional form of the volume exclusion term for residues that are not in contact in the native complex are given by a short range repulsive interaction, which is described elsewhere~\cite{Hyeon06Structure}. The other is the Coulomb potential~\cite{Lee2001}, which has the form
\begin{eqnarray}
{\cal H}_{C} = k_BT l_B {q_{Mg} q_{aa}\over r}
\end{eqnarray}
where $l_B$ is the Bjerrum length, $r$ is the  distance between the Mg$^{2+}$ ion and a residue, $q_{Mg} = +2$ is the electric charge or the ion, and $q_{aa}$ is the effective electric charge of a residue, which is set to be $-1$. Here, we assume that the electrostatic interaction between the E37 of the D1 domain of the ICAM-3 and the Mg$^{2+}$ ion is sensitive to the conformation of the I domain. It plays  a crucial in the binding of the I domain to the ICAM-3~\cite{GangSong2005,SpringerReview2007}. We have chosen the value 35nm for the Bjerrum length, following  a different study~\cite{Koculi2007} for RNA. For simplicity, we also choose the effective friction constant of the $Mg^{2+}$ ion to be the same as in our previous study~\cite{Hyeon06Structure}.%, which is for the monomers of the protein and assume that the net results of the DFS are insensitive to this parameter.

\bigskip
{\bf Acknowledgements}: We are grateful to Naoto Hori for useful discussions and for help with the figures. This work, which was completed when the authors were at the University of Maryland College Park, was supported by a grant from the National Institutes of Health (R01 GM089685). Additional support from the Collie-Welch Chair (F-0019) is gratefully acknowledged.
%\begin{figure}[h]
%  \centering
%  \includegraphics[bb = 0 0 3560 590, width = 4in]{1zon_2s}
%  \caption{Secondary structures of I domain (pdb code 1ZON).}\label{fig:1zon2s}
%\end{figure}
\newpage
\singlespace
%\bibliographystyle{pnas}
%\bibliography{integrin_bib_1,mybib1}

\setstretch{2.0}
\newpage
%\section*{Table captions}
\noindent{\bf Table 1~\ref{tab:gammak}}: Parameter values of $\gamma$ and $k^o$ obtained by fitting the two-linear regions in the [$f^*,\ln r_f]$ plot (Fig.~\ref{fig:forcerf}b) to the Bell-Evans model (Eq.~\ref{eqn:f*}).
%\newpage
\begin{table}
  \centering
   \begin{tabular}{|c|c|c|c|c|c|c|}
      \hline
      & Complex & $\gamma_1$(\AA) & $k_1^o$ $(s^{-1})$ & $\gamma_2$(\AA) & $k_2^o$ $(s^{-1})$\\
      \hline
      \multirow{2}{*}{Simulations}& LFA-1/ICAM-1 &  1.9 & 50.6& 5.3 &0.01\\
      & LFA-1/ICAM-3 &  3.8 & 45.3& 7.8 &0.12\\
      \hline
      \multirow{2}{*}{Experiment~\cite{Experiment2}}&LFA-1/ICAM-1 &  0.6 & 17 & 3.5 & 0.02\\
      &LFA-1/ICAM-2 &  1.5 & 13 & 4.9  & 0.06\\
      \hline
    \end{tabular}
  %\caption{}
  \label{tab:gammak}
\end{table}

\newpage
\section*{Figure captions}

{\bf Figure~\ref{fig:ID1}}: Cartoon representation of the $\alpha$L I domain $-$ ICAM3-D1 complex used in the simulations. The grey sphere is the $Mg^{2+}$ ion. To mimic the experimental setup we tethered several residues in the ICAM3-D1 domain to an immobilized surface through springs. Similarly residues in the $\alpha$L I domain are attached to a virtual plane, which is pulled at a constant velocity or at a constant force. 

\noindent{\bf Figure~\ref{fig:fpiezo}}: Typical force-spring-movement curves in the unbinding simulations. Force increases linearly till the complex ruptures around $f \approx$ 60pN. After the rupture the force precipitously drops to zero. This sample trajectory is generated at a constant loading rate.  

\noindent{\bf Figure~\ref{fig:forcerf}}: (a) Distribution of the unbinding forces at three loading rates obtained from SOP simulations for LFA-1/ICAM-1 complex. (b) Plot of the most probable force obtained from simulations as a function of the loading rate for LFA-1-ICAM-$x$ with $x$ = 1 and 3.  Results for LFA-1/ICAM-1 and LFA-1/ICAM-2 complexes are taken from~\cite{Experiment2}. Note that the change in slopes is significantly more prnounced in experiments than simulations for LFA-1/ICAM-1.

\noindent{\bf Figure~\ref{fig:gammarf}}: Dependence of $\gamma$, the location of the transition state, as a function of the loading rate, $r_f$ for two CACs. The results, obtained from simulations, show an abrupt change in $\gamma$ from a large to a small value over a narrow range of $r_f$. 

\noindent{\bf Figure~\ref{fig:pdf}}: (a) Distribution of $\Lambda$, the collective reaction coordinate (Eq.\ref{rxncor}) for LFA-1/ICAM-3. Interestingly, in this structure based reaction coordinate the CAC behaves as a two-state system. (b) Free energy profiles as a function of $\Lambda^{\prime} = \Lambda - \Lambda_{nat}$ ($\Lambda_{nat}$ is the value of $\Lambda$ in the native state) at various constant force values. The forces decrease from 50 pN (top curve) to 80 pN (bottom curve). The construction of the free energy profiles is described in the text.

\noindent{\bf Figure~\ref{fig:fep}}: (a), (b) and (c) Free energy profiles for the individual binding pairs at $F=$ 50, 67, and 80 pN, respectively. The averages are given in dashed lines. (d) Residue numbers of the binding pairs for which the free energy levels are given (a) - (c).

\newpage

\begin{figure}
   \centering
   \includegraphics[width = 4.5in]{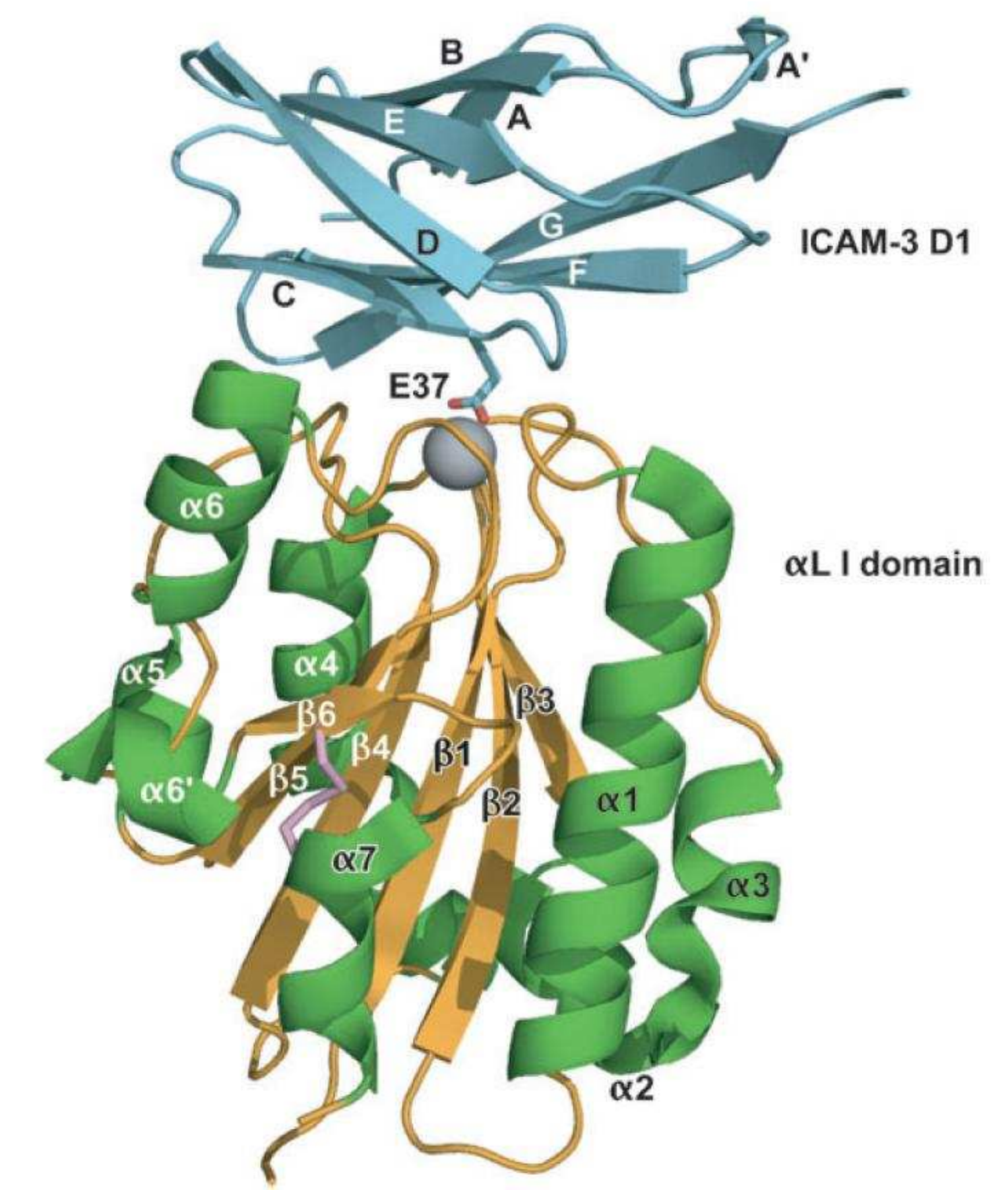}
   \caption{}\label{fig:ID1}
\end{figure}

\begin{figure}
   \centering
   \includegraphics[width = 4.5in]{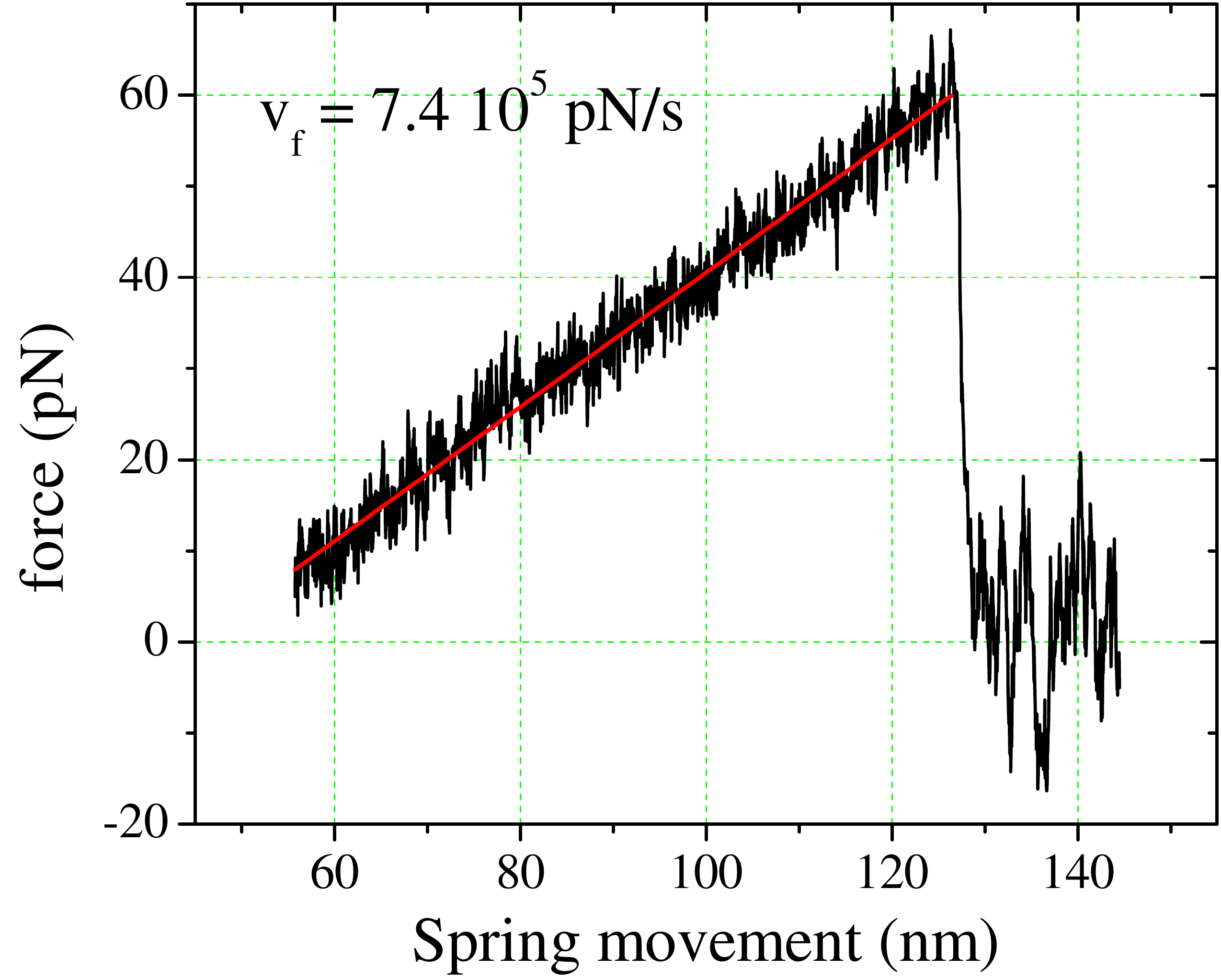}
   \caption{}\label{fig:fpiezo}
\end{figure}

\begin{figure}
   \centering
   \subfigure[]
   {
        \includegraphics[width = 4.5in]{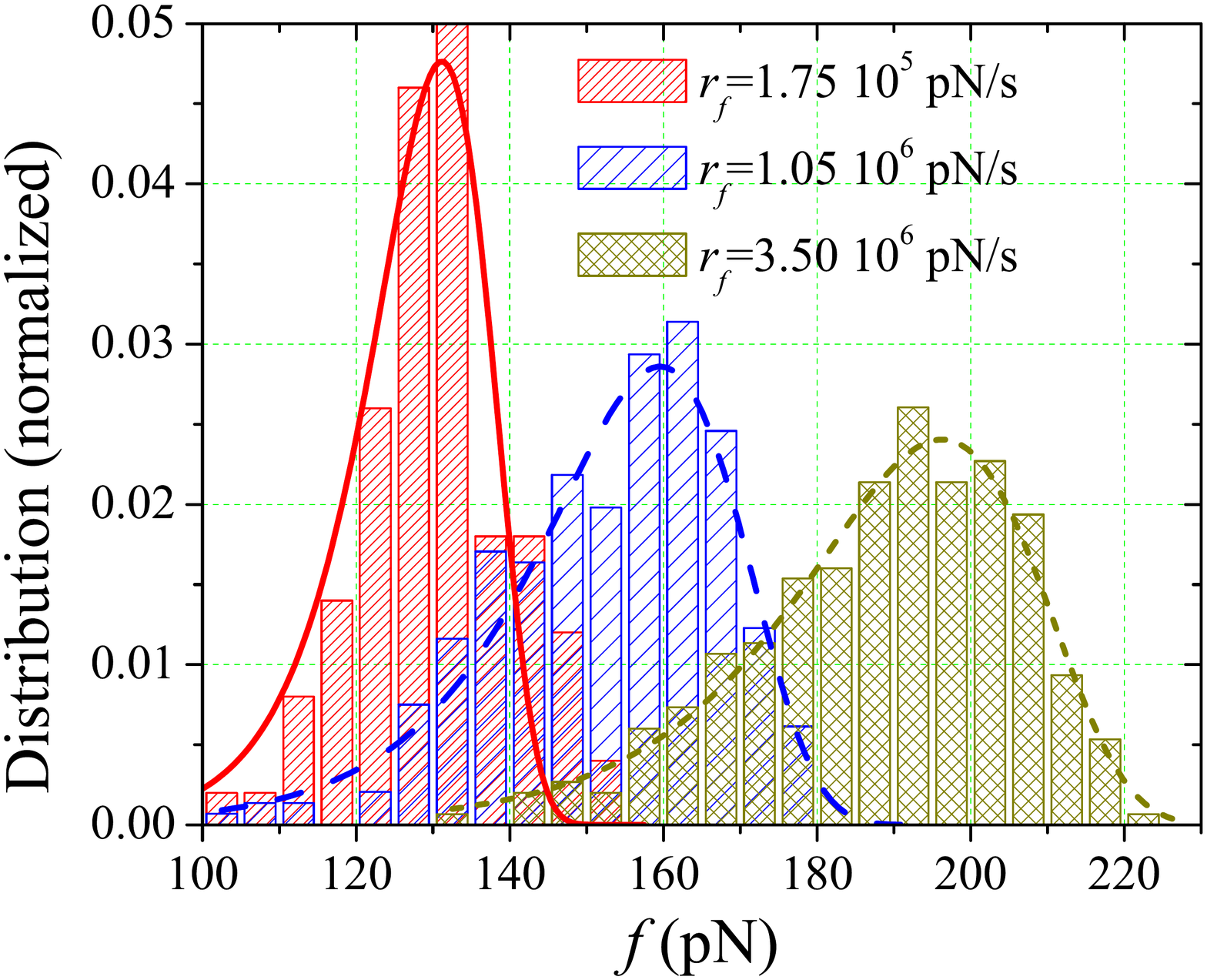}
        \label{subfig:forcePDF}
   }
   \subfigure[]
   {
        \includegraphics[width = 4.5in]{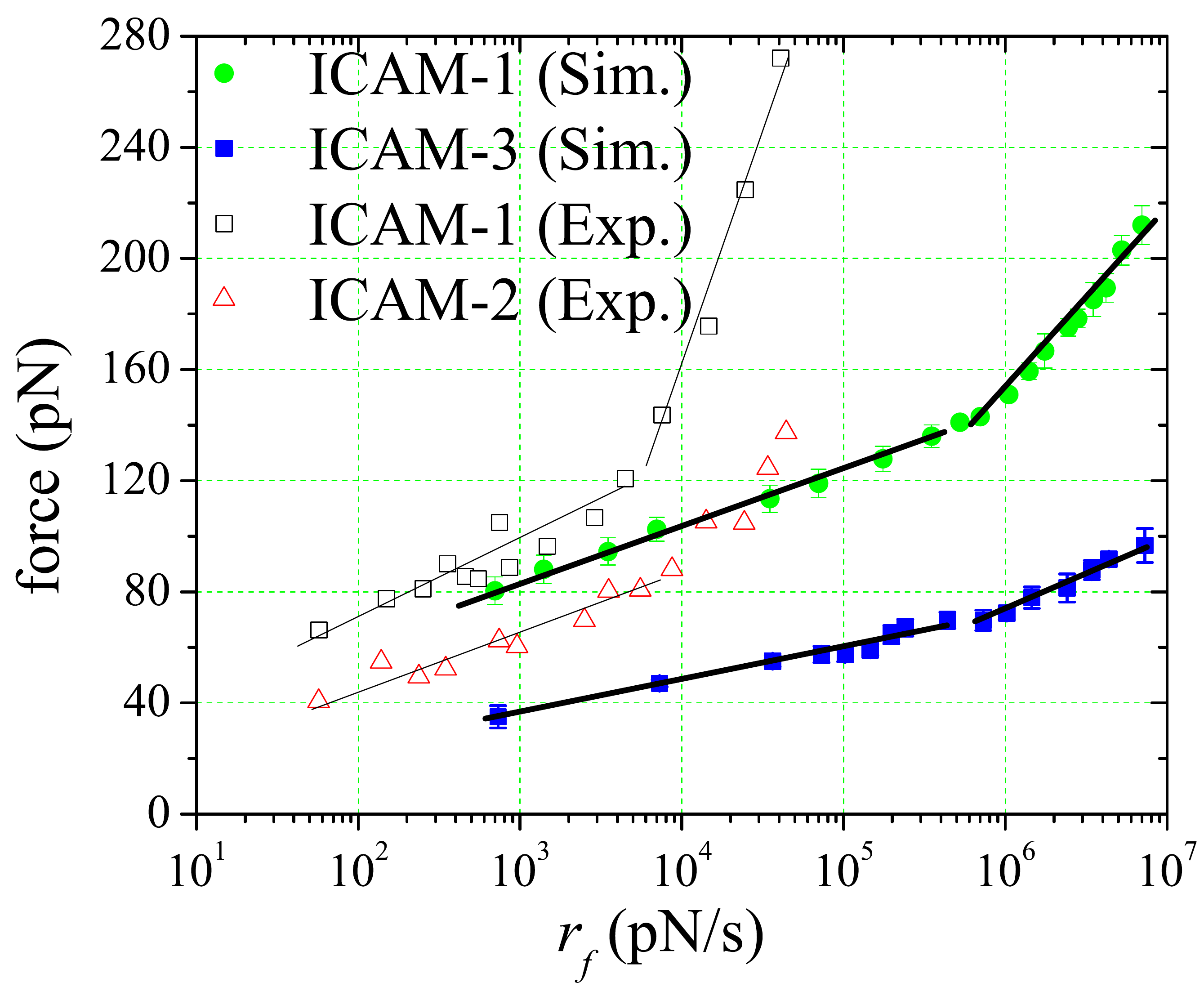}
        \label{subfig:forcerf}
   }
   \caption{}\label{fig:forcerf}
\end{figure}

\begin{figure}
   \centering
   \includegraphics[width = 4.5in]{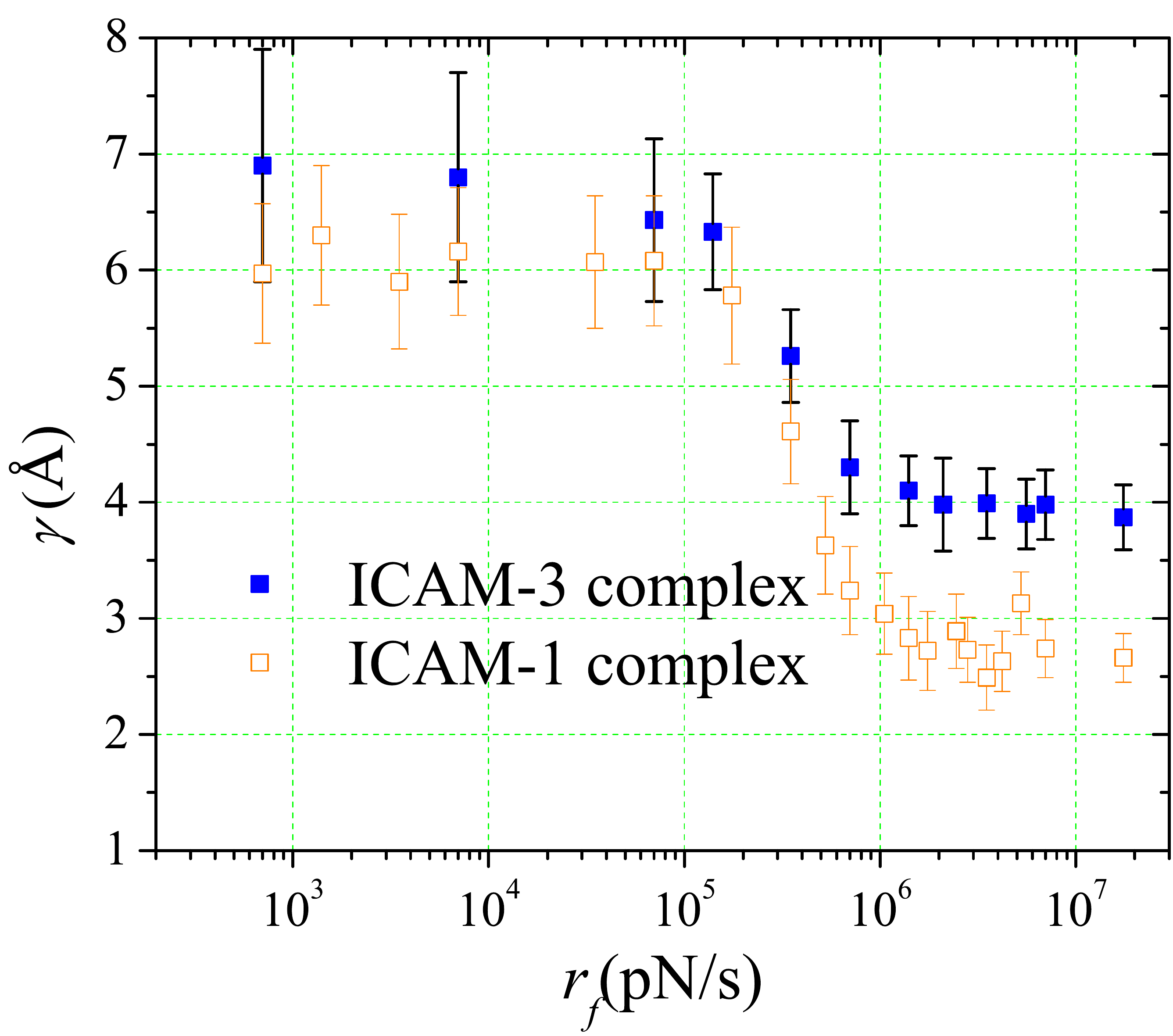}
   \caption{}\label{fig:gammarf}
\end{figure}

\begin{figure}
   \centering
   \subfigure[]
   {
        \includegraphics[width = 4.5in]{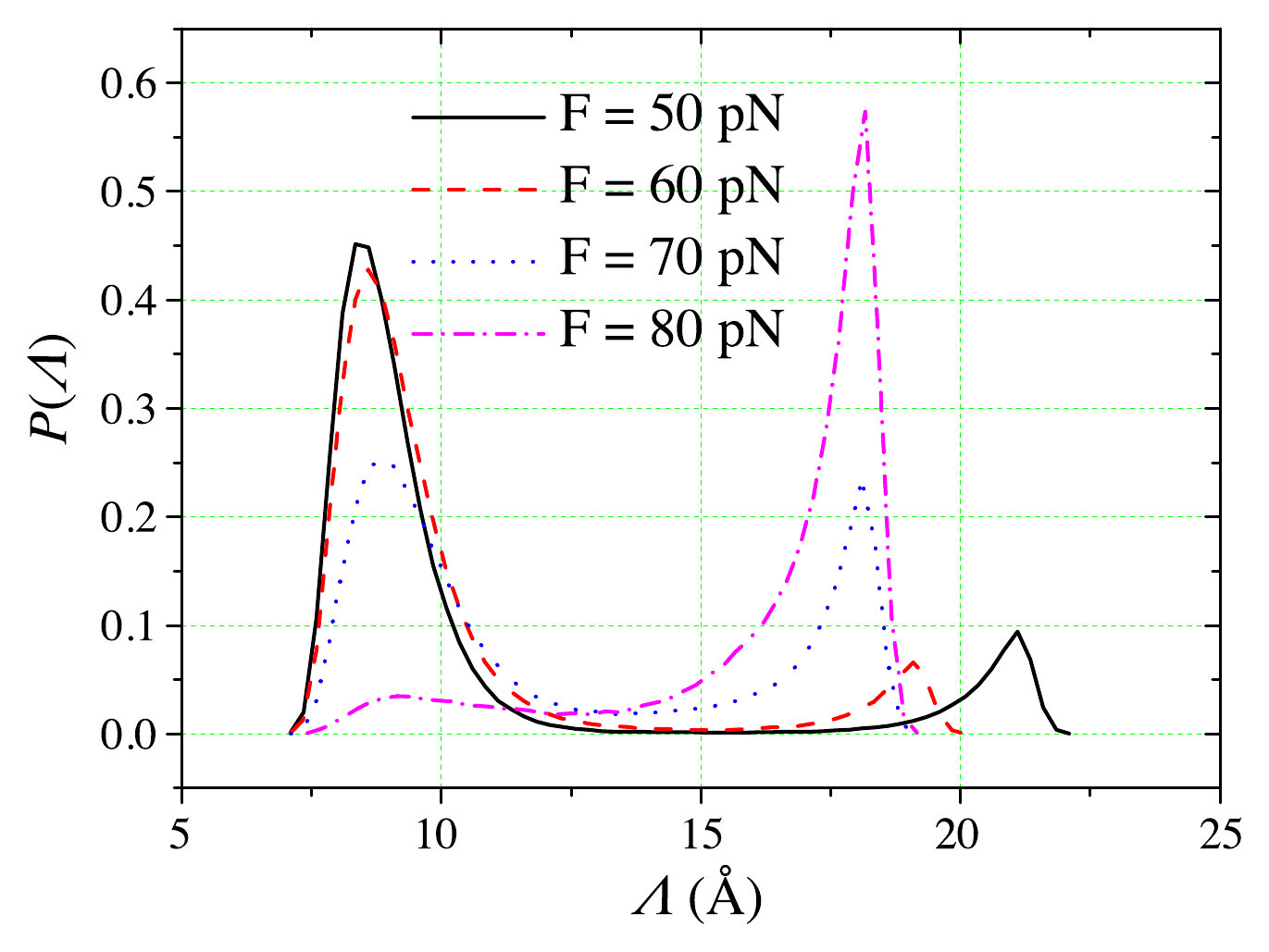}
        \label{subfig:pdf}
   }
   \subfigure[]
   {
        \includegraphics[width = 4.5in]{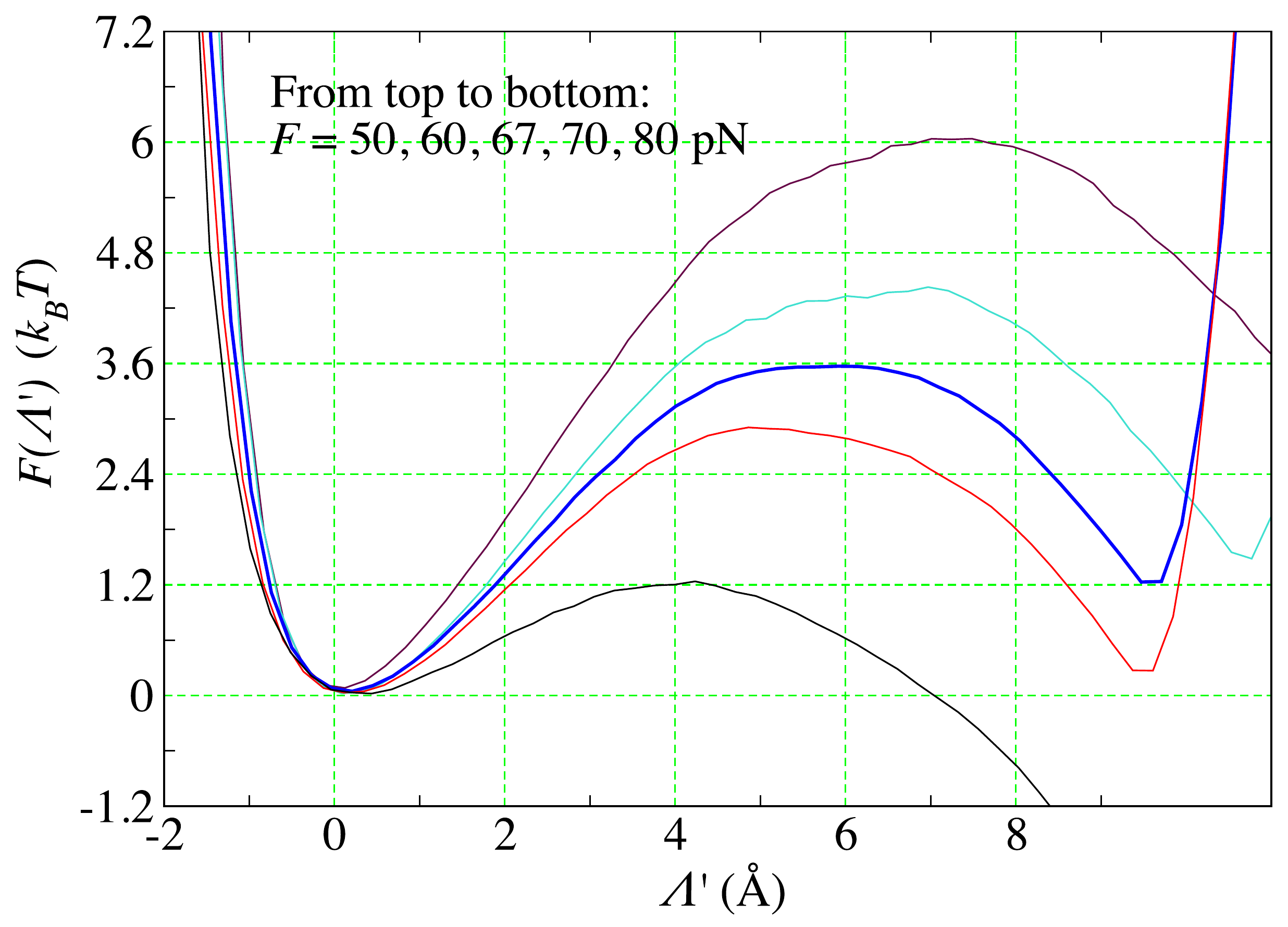}
        \label{subfig:FEPs}
   }
   \caption{}\label{fig:pdf}
\end{figure}

\begin{figure}
   \centering
   \subfigure[]
   {
        \includegraphics[width = 2.5in]{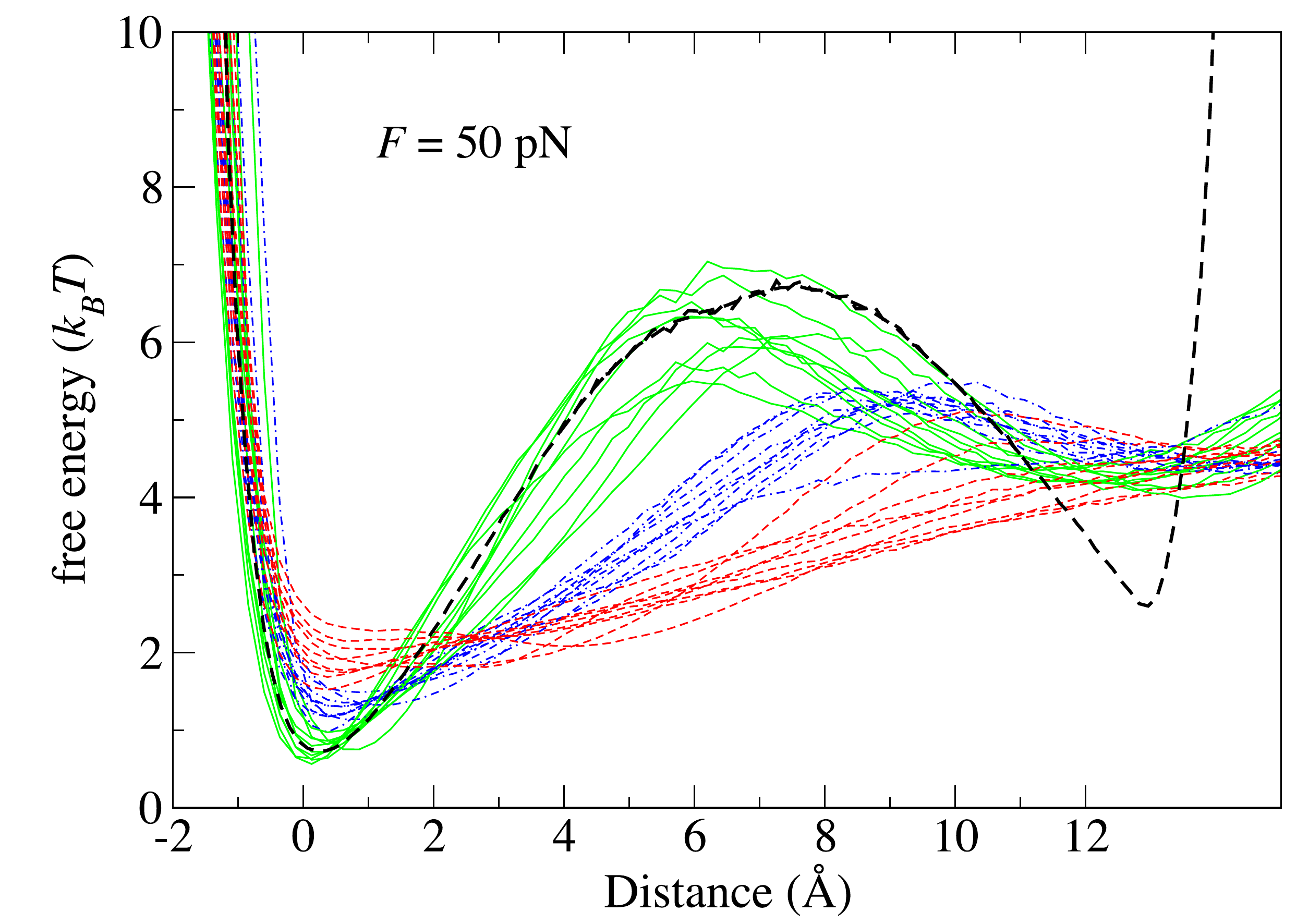}
   }
   \subfigure[]
   {
        \includegraphics[width = 2.5in]{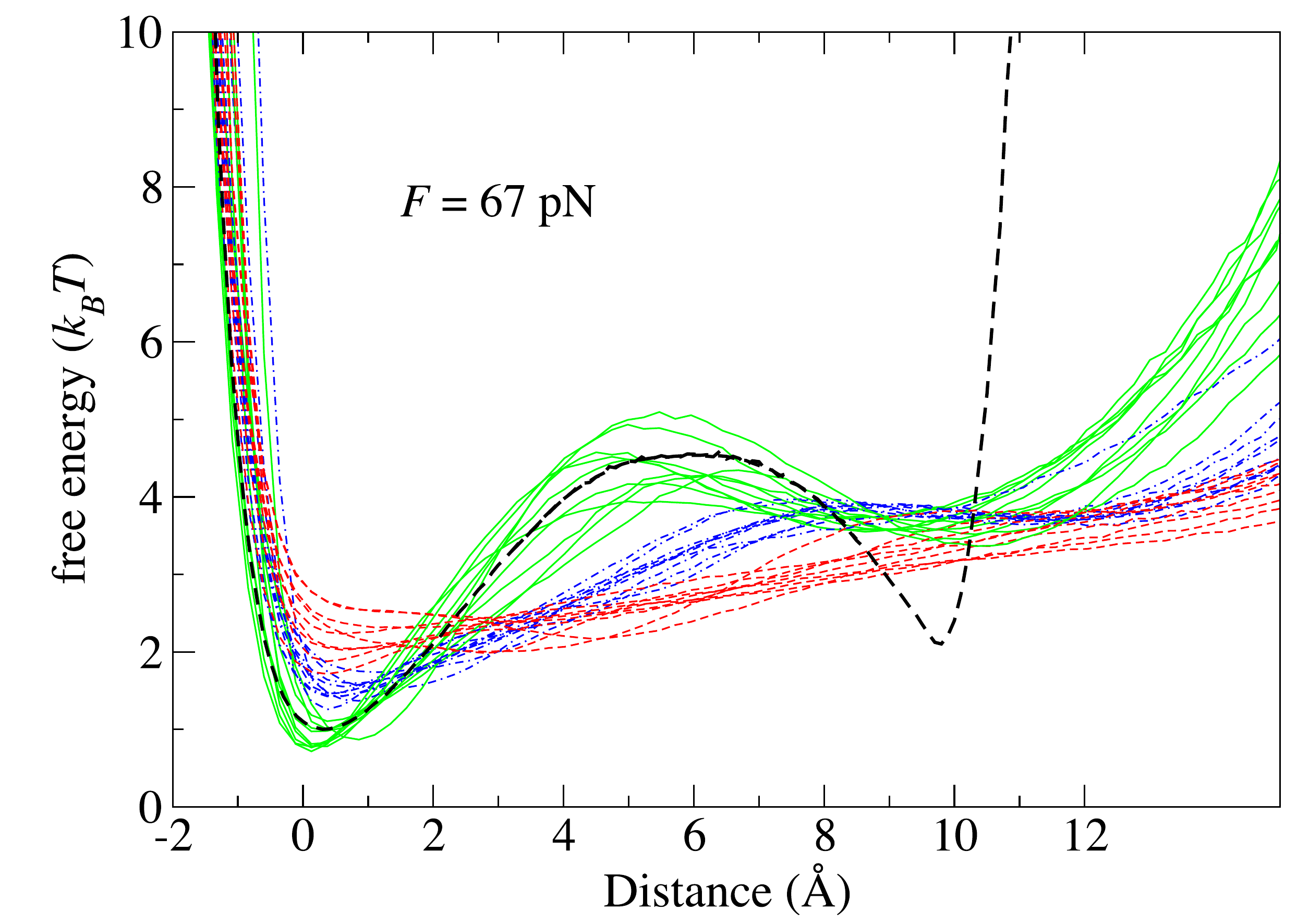}
   }\vspace{0.2in}\\
   \subfigure[]
   {
        \includegraphics[width = 2.5in]{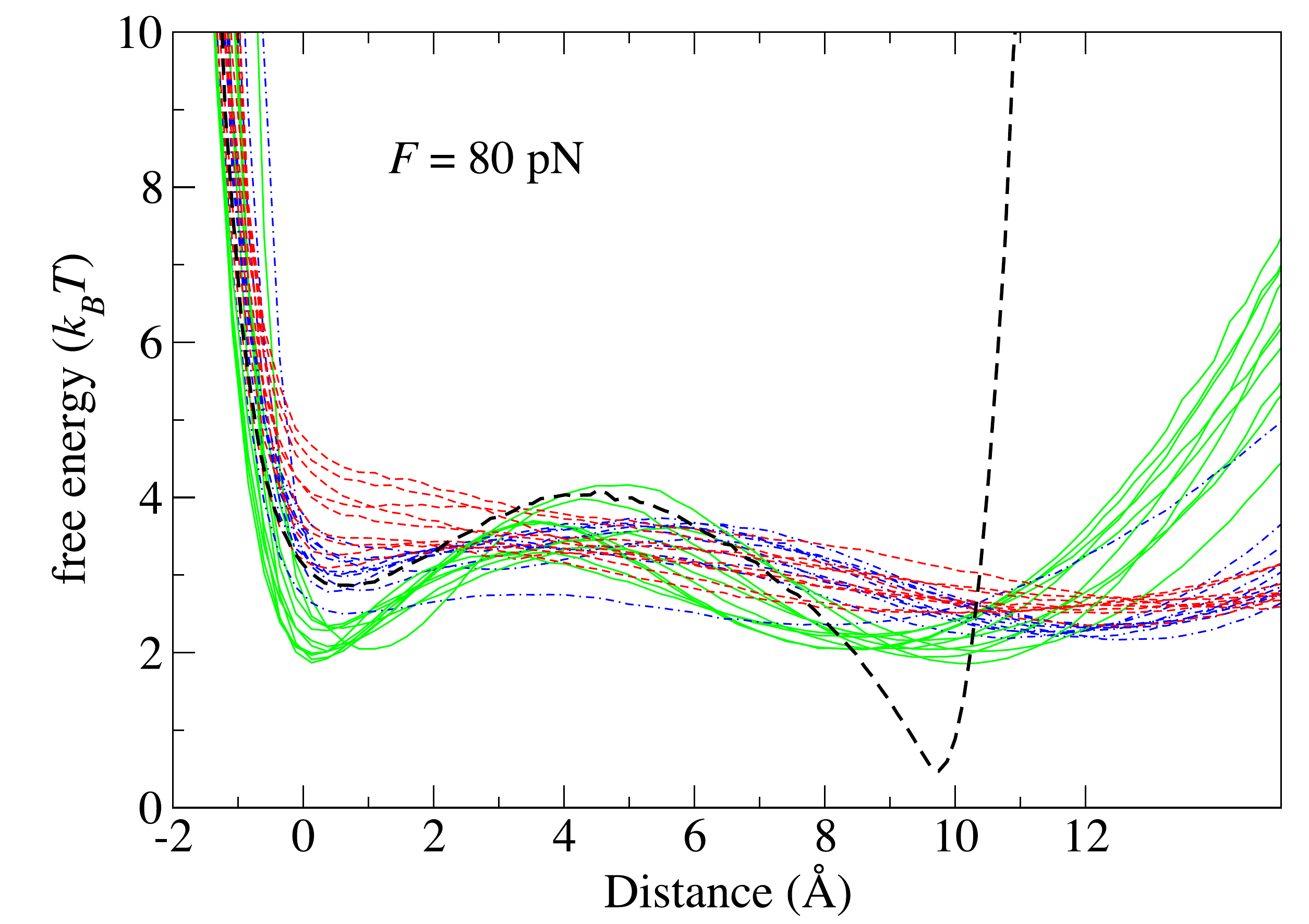}
   }
   \subfigure[]
   {
        \includegraphics[width = 2.5in]{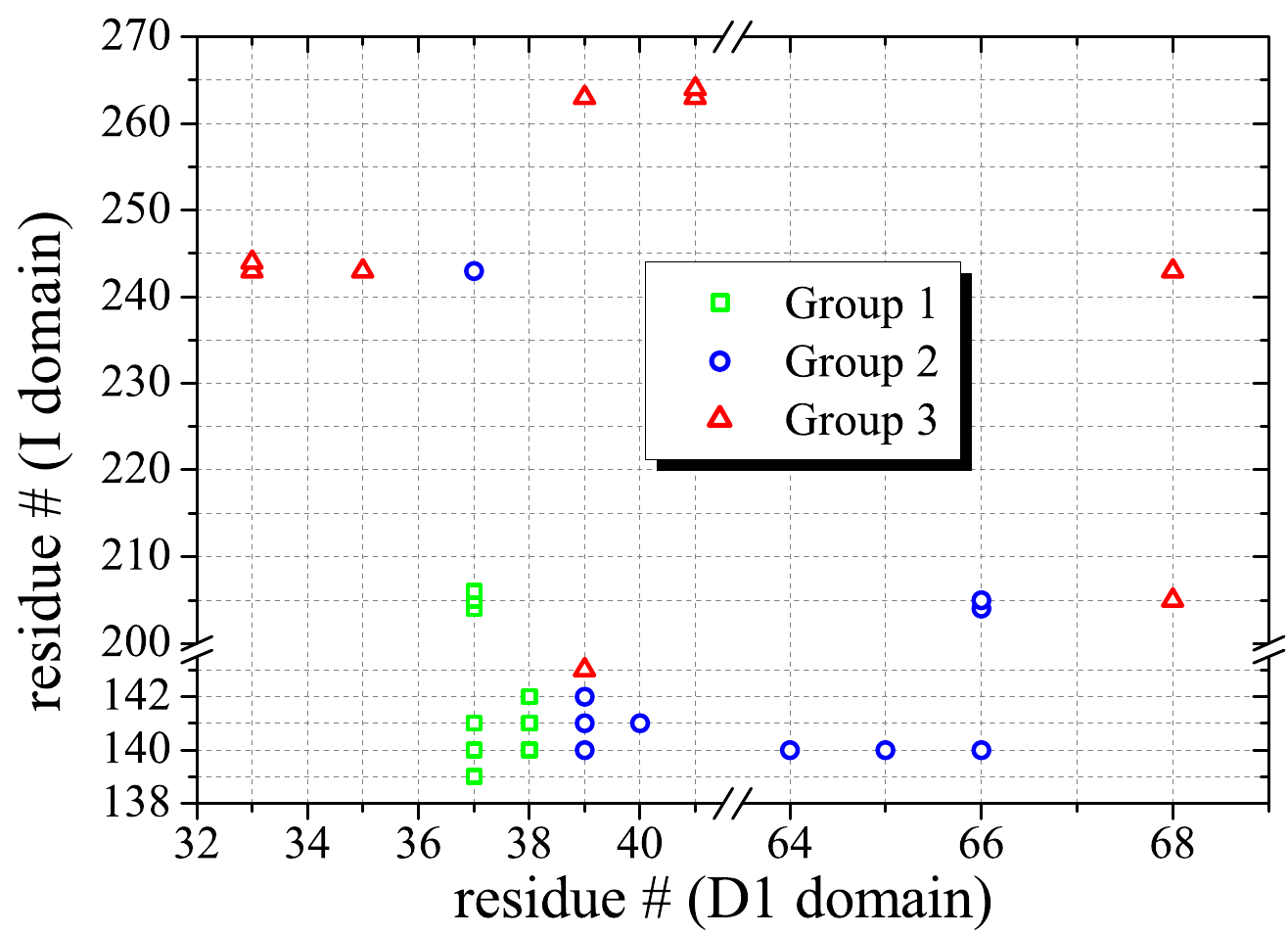}
   }
   \caption{}\label{fig:fep}
\end{figure}

\end{document}